\documentstyle[epsf,aps,12pt,preprint,amsfonts]{revtex}

\title{Statistics of Certain Models of Evolution}

\author{
Russell K. Standish}
\address{
High Performance Computing Support  Unit\\
University of New South Wales\\
Sydney, 2052\\
Australia\\
R.Standish@unsw.edu.au
}

\newcommand{\reals}{{\sf R\hspace*{-0.9ex}\rule{0.15ex}%
{1.5ex}\hspace*{0.9ex}}}

\newcommand{\br}{\mbox{\boldmath{$r$}}}          
\newcommand{\bbeta}{\mbox{\boldmath{$\beta$}}}   
\newcommand{\bmu}{\mbox{\boldmath{$\mu$}}}       
\newcommand{\bn}{\mbox{\boldmath{$n$}}}          
\newcommand{\nsp}{\mbox{$n_{\rm sp}$}}           

\tightenlines
\draft

\begin{document}
\maketitle

\begin{abstract}
  In a recent paper, Newman\cite{Newman97b} surveys the literature on
  power law spectra in evolution, self-organised criticality and
  presents a model of his own to arrive at a conclusion that
  self-organised criticality is not necessary for evolution. Not only
  did he miss a key model (Ecolab\cite{Standish94,Standish96}) that
  has a clear self-organised critical mechanism, but also Newman's
  model exhibits the same mechanism that gives rise to power
  law behaviour as does Ecolab. Newman's model is, in fact, a ``mean
  field'' approximation of a self-organised critical system.
  
  In this paper, I have also implemented Newman's model using the
  Ecolab software, {\em removing} the restriction that the number of
  species remains constant. It turns out that the requirement of
  constant species number is non-trivial, leading to a global coupling
  between species that is similar in effect to the species
  interactions seen in Ecolab. In fact, the model must self-organise
  to a state where the long time average of speciations  balances
  that of the extinctions, otherwise the system either collapses or explodes.
  
In view of this, Newman's model does not provide the hoped-for counter
example to the presence of self-organised criticality in evolution,
but does provide a simple, almost analytic model that can used to
understand more intricate models such as Ecolab.

\end{abstract}
\narrowtext

\section{Introduction}

Over the last five years, the notion that Biological Evolution is a
{\em self-organised critical phenomenon} has gained currency, and in
particular, has been championed by Bak\cite{Bak-Sneppen93} and
Kauffman\cite{Kauffman93}. Self-organised critical phenomena are
characterised by a frustration between two processes. The archetypical
example is that of a sandpile, where the process of adding sand to a
sand pile makes the slope of that pile steeper is opposed by the
instability of the sandpile which works to make the sandpile flatter
once the slope passes a critical angle. One of the the most obvious
manifestations of criticality is a power law spectral behaviour,
although criticality is by no means necessary for this power law
behaviour to be manifest. 

In a recent paper, Newman\cite{Newman97b} surveys the field to conclude
that the mechanism by which ecosystems are driven to criticality is
not well understood, but that the evidence in the fossil record for
power law spectra of extinction event size and species lifetimes is
good.  Sol\'e et.  al.\cite{Sole-etal97} present the best evidence yet
that these distributions are power laws. However, Newman missed an
important model of evolution, {\em
  Ecolab}\cite{Standish94,Standish96}, that is more general than those
surveyed, and gives us the best idea yet of how evolution could be a
self-organised critical phenomenon.

Newman goes further to introduce his own model of evolution to make
the point that the coevolutionary avalanches that all the other models
(including Ecolab) exhibit are not necessary for the observed power law
behaviour. He further claims that his model is not critically
self-organised. However, the mechanism that leads to power law
behaviour in Newman's model is precisely the same as that in Ecolab,
and that mechanism is of the nature of a frustration between two
processes that characterises Bak's sandpile model. 

\section{Ecolab}

In this section, we consider a model of evolution called {\em
  Ecolab}. Ecolab (perhaps unfortunately) is both the name of a model
  and a simulation system written by the author to implement that model.
The ecology is described by a generalised Lotka-Volterra equation,
which is perhaps the simplest ecological model to use.
\begin{equation} \label{lotka-volterra}
\dot{n}_i = r_i n_i + \sum_{j=1}^{\nsp}\beta_{ij}n_in_j
\end{equation}
Here \br{} is the difference between the birth rate and death rate for
each species, in the absence of competition or symbiosis. \bbeta{} is
the interaction term between species, with the diagonal terms
referring to the species' self limitation, which is related in a
simple way to the carrying capacity $K_i$ for that species in the
environment by $K_i=-r_i\beta_{ii}$.  In the literature (eg
Strobeck\cite{Strobeck73}, Case\cite{Case91}) the interaction terms
are expressed in a normalised form, $\alpha_{ij}=-K_i/r_i\beta_{ij}$, and
$\alpha_{ii}=1$ by definition. \bn{} is the species density.

These equations are simulated on a simulator called {\em
  Ecolab}.\cite{Ecolab-Tech-Report} The vectors \bn{} and \br{} are stored
as dynamic arrays, the size of which (i.e. the system dimension) can
change in time. 

\subsection{Linear Stability Analysis}

Linear analysis starts with the fixed point of equation (\ref{lotka-volterra})
\begin{equation}\label{fixed point}
\hat{\bn} = -\bbeta^{-1}\br,
\end{equation}
where $\dot{\bn}=0$. There is precisely one fixed point in the
interior of the space of population densities (i.e. \bn{} such that
$n_i>0$) provided that all components of $\hat{\bn}$ are
positive, giving rise to the following inequalities:
\begin{equation}\label{positive species}
\hat n_i = \left(\bbeta^{-1}\br\right)_i>0,\;\; \forall i
\end{equation}
This interior space is denoted $\reals_+^{\nsp}$ mathematically.

There may also be fixed points on the boundary of $\reals_+^{\nsp}$,
where one or more components of \bn{} are zero (corresponding to an
extinct species). This is because the subecology with the living
species only (i.e. with the extinct species removed) is equivalent to
the full system.

The {\em stability} of this point is related to the
negative definiteness of derivative of $\dot{\bn}$ at $\hat{\bn}$. The
components of the derivative are given by
\begin{equation}\label{derivative}
\frac{\partial\dot{n}_i}{\partial n_j} =
\delta_{ij}\left(r_i+\sum_k\beta_{ik}n_k\right) + \beta_{ij}n_i
\end{equation}
Substituting eq (\ref{fixed point}) gives
\begin{equation}
\left.\frac{\partial\dot{n}_i}{\partial n_j}\right|_{\hat{\bn}}=
-\beta_{ij}\left(\bbeta^{-1}\br\right)_i
\end{equation}

Stability of the fixed point requires that this matrix should be
negative definite. Since the $\left(\bbeta^{-1}\br\right)_i$ are
all negative by virtue of (\ref{positive species}), this is equivalent
to \bbeta\ being negative definite, or equivalently, that its \nsp\
eigenvalues all have negative real part. Taken together with the
inequalities (\ref{positive species}), this implies that $2\nsp$
inequalities must be satisfied for the fixed point to be stable. This
point was made by Strobeck\cite{Strobeck73}, in a slightly different
form. (Note that Strobeck implicitly assumes that
$\sum_ir_i\hat{n}_i / K_i >0$, so comes to the conclusion that
$2\nsp-1$ conditions are required.)  If one were to randomly pick
coefficients for a Lotka-Volterra system, then it has a probability of
$4^{-\nsp}$ of being stable, i.e. one expects ecosystems to become
more unstable as the number of species increases\cite{May74}.

\subsection{Permanence}

Whilst stability is a nice mathematical property, it has rather less
relevance when it comes to real ecologies. For example the traditional
predator-prey system studied by Lotka and Volterra has a limit
cycle. The fixed point is decidedly unstable, yet the ecology is {\em
permanent} in the sense that both species' densities are larger than
some threshhold value for all time. Hofbauer et
al. \cite{Hofbauer-etal87} and Law and Blackford\cite{Law-Blackford92}
discuss the concept of {\em permanence} in Lotka-Volterra systems,
which is the property that there is a compact absorbing set ${\cal
M}\subset\reals^{\nsp}_+$ {\em i.e} once a trajectory of
the system has entered ${\cal M}$, it remains in ${\cal M}$. They derive a sufficient
condition for permanence due to Jansen\cite{Jansen87} of the form:
\begin{equation}\label{Jansen}
\sum_ip_if_i(\hat{\bn}_B) =
\sum_ip_i(r_i-\sum_j\beta_{ij}\hat{n}_{Bj}) > 0, \;\; \exists p_i>0 
\end{equation}
for every $\hat{\bn}_B$ equilibrium points lying on the boundary
($\hat{n}_{Bi}=0 \;\; \exists i$), provided the system is {\em
  bounded} (or equivalently {\em dissipative}).\footnote{Boundedness
  is ensured in this model by choosing the $\beta_{ij}$ such that
  $\beta_{ij}+\beta_{ji}\leq0, \,\,\forall i,j$. This precludes
  symbiosis, but does allow for unstable behaviour. See
  \cite{Ecolab-Tech-Report} for a discussion of boundedness} This
condition is more general than stability of the equilibrium --- the
latter condition implies that a local neighbourhood of the equilibrium
is an absorbing set. Also, the averaging property of Lotka-Volterra
systems implies that the equilibrium must lie in the positive cone
$\reals^{\nsp}_+$.  So (\ref{positive species}) must still hold for
permanence.

Consider the boundary points $\hat{\bn}_B$ that are
missing a single species $i$. Then Jansen's condition for these
boundary points is
\begin{equation}\label{single-deficiency}
r_i-\sum_j\beta_{ij}\hat{n}_{Bj}>0.
\end{equation}
This set of conditions is linearly independent. Let the number of such
boundary points be denoted by $n_B\leq\nsp$. Then the set of
conditions (\ref{Jansen}) will have rank $n_B\leq\nu\leq\nsp$ (the
number of linearly independent conditions, so the system has at
most probability $2^{-\nsp-\nu}$ of satisfying Jansen's  permanence condition
if the coefficients are chosen uniformly at random. As stability is
also sufficient for permanence, the probability lies between
$4^{-\nsp}$ and $2^{-\nsp-\nu}$.

Another rather important property is {\em resistance to
invasion}.\cite{Case91} Consider a boundary equilibrium
$\hat{\bn}_B$. If it is proof against invasion from the missing
species, then the full system cannot be permanent. For the boundary
points that miss a single species, this implies that condition
(\ref{single-deficiency}) is necessarily satisfied for permanence,
along with  (\ref{positive species}). The probability of
permanence is then bounded above by $2^{-\nsp-n_B}$.

The important point to take away from this section is that whilst a
randomly selected ecology is more likely to be permanent than to have
a stable equilibrium, the likelihood decreases exponentially with
increase in species number.

\subsection{Mutation}

Adding mutation involves adding an additional operator to
equation (\ref{lotka-volterra}) 
\begin{equation}
\dot{\bn} = \br*\bn + \bn*\bbeta\bn + {\tt mutate}(\bmu,\br,\bn)
\end{equation}
where $*$ refers to elementwise multiplication. This operator extends
the dimension of the whole system, so is rather unusual. It is not
germane to the present argument what the precise form of {\tt mutate}
is, the interested reader is referred to the previous publications
describing it \cite{Standish94,Standish96,Ecolab-Tech-Report}. Suffice
it to say, that it adds new species according to a stochastic
mechanism, and that we would expect that the criticality result to be
robust with respect to changes of mutation algorithm employed.

\subsection{Self Organised Criticality}

Lets consider what happens to the largest eigenvalue of \bbeta.
Suppose initially, the system has a stable equilibrium, in which case
all the eigenvalues have negative real part. As mutations are added
to the system, the largest eigenvalue will increase towards zero. As
it passes zero, the system destabilises, and the system will start to
exhibit limit cycles or chaotic behaviour. As further mutations are
added to the system, permanence is no longer satisfied, and an extinction
event will occur. This will restore permanency to the system, and possibly
even stability. So we have two frustrated processes opposed to each
other, the first, mutation, which builds up ecosystem complexity, and
the second being the trend to impermanency as ecosystem become more
complex. This is analogous to the sand being added to the top of the
pile, and the stability of the sandpile slope in Bak's sandpile model.

\section{The Newman model}

Newman has presented his model of evolution in a number of
papers\cite{Newman96,Newman97a,Newman97b}, and is largely equivalent
to an earthquake model presented in
\cite{Newman-Sneppen96,Sneppen-Newman97}. In the biological context,
the model has a fixed number of species, all of which feel an
environmental stress, denoted by $\eta(t)$, which is random variate
with distribution $p_{\rm stress}(\eta)$. Each species has an
individual threshold $x_i$, such that if $\eta>x_i$, species $i$
becomes extinct. These extinct species are then replaced by new
species, with thresholds randomly assigned from some distribution
$p_{\rm thresh}(x)$. There is one further twist to the model, in that
the threshold values are allowed to drift over time in order to
prevent the model stagnating with every species having the maximum threshold.

The Ecolab software allows us to build a variant of this model that
allows the number of species to vary over time. When the model was
first implemented, the system underwent a ``mutation catastrophe'', in
which the number of species exploded, This is similar to what happens
in the Ecolab model when the mutation rate is set too high. Normally,
one would expect that the number of speciation events should be
proportional to the number of species. However, this leads to an
excess of speciation over extinctions.

The resolution of this conundrum is to require that the stress values
$\eta$ be proportional to the number of species, i.e.
$\eta=\nsp\eta'$, where $\eta'$ is drawn from some distribution
$p_{\rm stress}(\eta')$. The justification for making this assumption
can be seen by considering a simplified model of Ecolab (called
Ecolab\verb|--|), described in the next section. Of course, in
Newman's original model, \nsp\ is a constant, and so his model is
consistent with this modification.

Wilke and Martinetz\cite{Wilke-Martinetz97} examined a similar model,
in which they label the mutation rate $g$, and consider finite $f$,
rather than $f=0$ as I do here. They too note the conundrum of
exponential growth in species number, and resolve it by introducing an
arbitrary logistic constraint. My argument is that the reason for this
logistic constraint is that species must interact with each other, and
the greater the number and strengths of these interactions, the
greater the stresses are that are felt by the ecosystem.

It could be argued that the {\em raison d'\^etre} of the Newman model
is to study the effect of coherent extinction through exogenous
causes. However, these will always give rise to stress distributions
that are independent of species number. However, the stress distribution
will ultimately be dominated by the term that does depend on the
species number.

\begin{figure}
\epsfbox{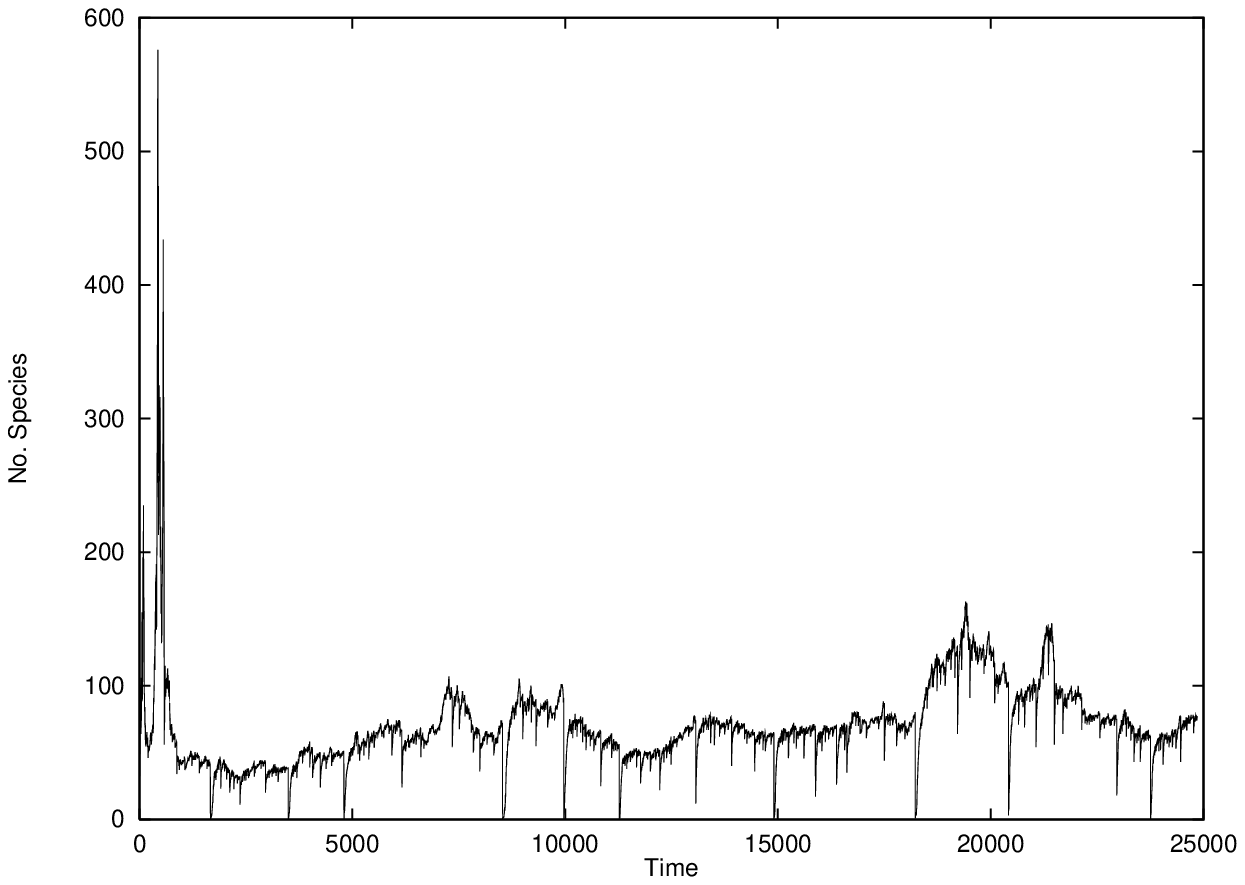}
\caption{\nsp{} as a function of time in the genralised Newman model.}
\label{newman-nsp}
\end{figure}

\begin{figure}
\epsfbox{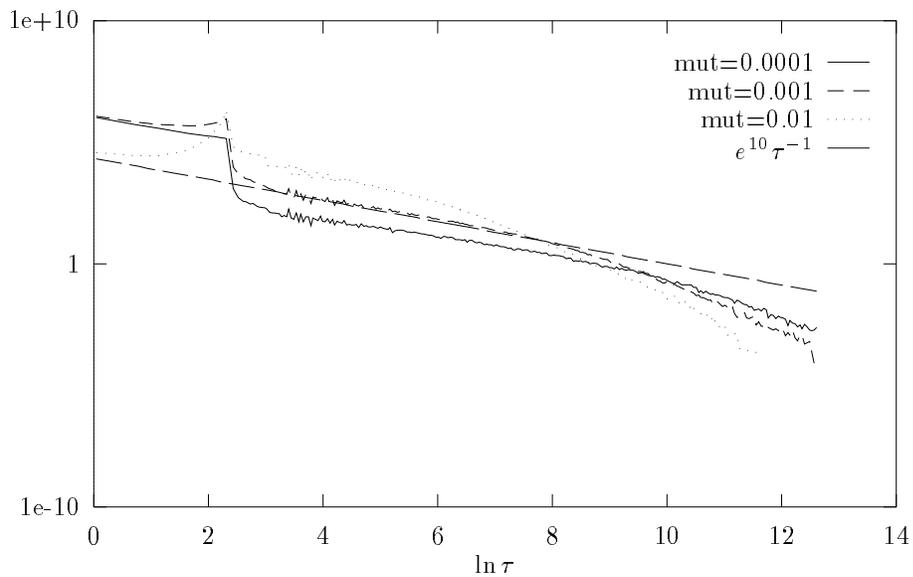}
\caption{Distribution of species lifetimes in generalised Newman model
with Gaussian stress distribution.}
\label{newman-lifetimes}
\end{figure}

Once the stress values depend on species number, the system
self-organises so that speciations and extinctions balance on
average. A trace of \nsp\ can be seen in Figure \ref{newman-nsp}, and
the distribution of lifetimes is seen in Figure
\ref{newman-lifetimes}. The peak in the curve at $\tau=10$ is an
artifact of the simulation, and should be ignored. The distribution
actually has two regions, the inner one $10\ll\tau\ll10^3$ having a
power law with exponent $\approx-1$, and the outer region
$\tau\gg10^3$ having exponent $\approx-2$. By running the experiment
at different mutation rates, the lifetime $\lambda$ at which the
distribution changed from $\tau^{-1}$ to $\tau^{-2}$ was found to be
inversely proportional to the mutation rate.

In comparing the result of my variation with the original Newman
model, it should be noted that the power law exponent in Newman's
original model is $-1$ out to a time $1/f$, and decays exponentially
after that. In my version, the same power law exponent was observed
out to $1/g$, and then appears to change to a faster power law decay,
although the error bars are sufficiently large not to rule out an
exponential decay. In each of these models, the lifetime $1/f$ or
$1/g$ respectively is roughly the lifetime that a maximally fit
organism (one with a maximal value $x_i$) can survive before sucumbing
to mutation pressures.

\section{The Ecolab\protect\verb|--| model}

In this section, we will consider a simplification of the Ecolab model
where the interaction terms $\sum_{j}\beta_{ij}n_j$ are replaced by a
random variate $\eta_i(t)$ from a suitable distribution:
\begin{equation}\label{ecolab--}
\dot n_i = (r_i - \eta_i)n_i
\end{equation}
Since $\eta_i$ is effectively a sum of a large number of independent
quantities, its distribution will tend to be normal, and the deviation
(controlling how large $\eta_i$ gets) will be proportional to \nsp,
the connectance (proportion of nonzero elements in \bbeta) and the
interaction strength.  This is why stresses in the Newman model must
be proportional to \nsp.  When $\eta_i$ exceeds $r_i$ for any
significant period of time, species $i$ becomes extinct. Since
$\eta_i(t)$ is a continuous function of $\bn(t)$ which is itself a
continuous function of $t$, there will be a correlation
$\eta(t)\eta(t+\tau)>0, \forall\tau<\tau_0, \exists\tau_0>0$. Equation
(\ref{ecolab--}) connects the full Ecolab model with the Newman model.

In order to make the analysis simpler, we assume that $n_i$ are real valued,
rather than integers as in Ecolab. In order to detect when extinction
happens, we take an arbitrary threshold $\sigma$, such that if
$n_i<\sigma$, species $i$ is extinct.

\section{Distribution of Species Lifetimes}

Figure \ref{newman-lifetimes} shows the distribution of species
lifetimes (time from speciation to extinction) in the augmented Newman
model. This figure is not normalised, as a power law $x^\alpha$ has an
infinite integral. So the abcissa of the graph is not significant, but
the slope is. The lines are fitted by linear regression. Authors often
quote a correlation coefficient, however this is generally meaningless
on a log-log plot. Even the value of the slope is meant to be an
indication only, as the large relative error at high lifetime values can
lead to significant errors in the computed slope. 

\begin{figure}
\epsfbox{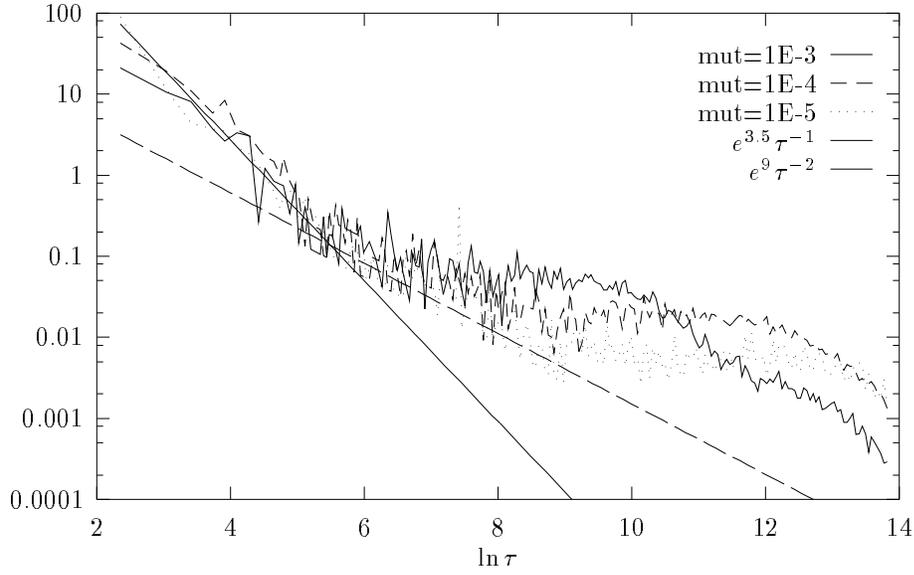}
\caption{Distribution of species lifetimes in Ecolab.}
\label{log-log}
\end{figure}

Figure \ref{log-log} shows the lifetime distribution for Ecolab which
has a slope of $-2$ for lifetimes less than 100, but $-1$ for larger
lifetimes. At still larger times ($\tau\gg 0.1/\mu$), the distribution
turns over, decaying exponentially. Previously published versions of
this graph \cite{Standish96} only show the smaller lifetime behaviour.

Consider now the probability $p(>\tau|x)$ that a species with
threshold $x$ will become extinct after time $t=\tau$ in the Newman
model. Since time is discrete in this model, this is simply the
probability that the stress $\eta$ does not exceed $x$ for the first
$\tau$ steps:
\begin{equation}
p(>\tau|x) = \left[\int_0^x p_{\rm stress}(\eta)d\eta\right]^{\tau}.
\end{equation}

Now the distribution $p(>\tau)$ of species having lifetimes $\tau$ is
just the above quantity, integrated over the distribution of
thresholds:
\begin{eqnarray}\label{p(.gt.tau)}
p(>\tau)&=&\int p(x) p(>\tau|x) dx\nonumber\\
&=& \int p_{\rm thresh}(x)
  \left[\int_0^x p_{\rm stress}(\eta)d\eta\right]^{\tau}
 \nonumber\\
&=& \int_0^1 p_{\rm thresh}(x) \xi^{\tau}\frac{dx}{d\xi}d\xi
\end{eqnarray}
where $\xi=\int_0^x p_{\rm stress}(\eta)d\eta$

Assume the following inequalities hold:
\begin{eqnarray}\label{ineq}
p_{\rm thresh}(x) &\leq & K_1 p_{\rm stress}(x), \,\, \forall x \nonumber\\
                  &\geq & K_0 p_{\rm stress}(x) \,\, \forall x<x_c, 
			\exists x_c
\end{eqnarray}
Without loss of generality, $p_{\rm thresh}(x)$ is taken to be the
uniform distribution between 0 and 1, and is zero outside this
interval. $p_{\rm stress}(x)$ is positive for all positive $x$, with
the large $x$ tail needed to establish power law
behaviour\cite{Newman97a}.  In this case, the constants $K_0$ and
$K_1$ correspond to the inverses of the maximum and minimum of $p_{\rm
thresh}(x)$ over the unit interval, and $x_c=1$. Let us introduce
$\xi_c=\int_0^{x_c}p_{\rm stress}(x)dx$ as being the change of
variable equivalent of $x_c$. In the case of uniform threshold
distribution, and monotonic stress distribution, $1-\xi_c$ is the
proportion of stress events that overwhelm the hardiest of
species. The inverse of this proportion is a time scale, above which
the lifetime distribution must decay exponentially. In order to
observe power law behaviour, the stress distribution must be chosen so
that $\xi_c\approx 1$.

Substituting eq (\ref{ineq}) into (\ref{p(.gt.tau)}) generates the
following inequality:
\begin{eqnarray}
K_0\int_0^{\xi_c}p_{\rm stress}(x)\xi^\tau \frac{dx}{d\xi}d\xi
&\leq p(>\tau) \leq &
K_1\int_0^1p_{\rm stress}(x)\xi^\tau \frac{dx}{d\xi}d\xi \nonumber\\
K_0 \frac{\xi_c^{\tau+1}}{\tau+1} &\leq p(>\tau) \leq & K_1
\frac1{\tau+1},
\end{eqnarray}
since $p_{\rm stress}(x)={d\xi}/{dx}$ and where
$\xi_c=\int_0^{x_c}p_{\rm stress}(x)dx$. 

Now $p(\tau)=p(>\tau-1)-p(>\tau)$, so the following inequality is
obtained:
\begin{equation}
\frac{(K_0\xi_c^\tau-K_1)\tau+K_0\xi_c^\tau}{\tau(\tau+1)} \leq p(\tau) \leq
\frac{(K_1-K_0\xi_c^\tau)\tau+K_1}{\tau(\tau+1)}
\end{equation}
Assuming that $\tau\ll(1-\xi_c)^{-1}$,
$\xi_c^\tau=(1+\tau(1-\xi_c)+\cdots\approx1$, this inequality may be
simplified:
\begin{equation}
\frac{(K_0-K_1)\tau+K_0}{\tau(\tau+1)} \leq p(\tau) \leq
\frac{(K_1-K_0)\tau+K_1}{\tau(\tau+1)}
\end{equation}

\begin{figure}
\epsfbox{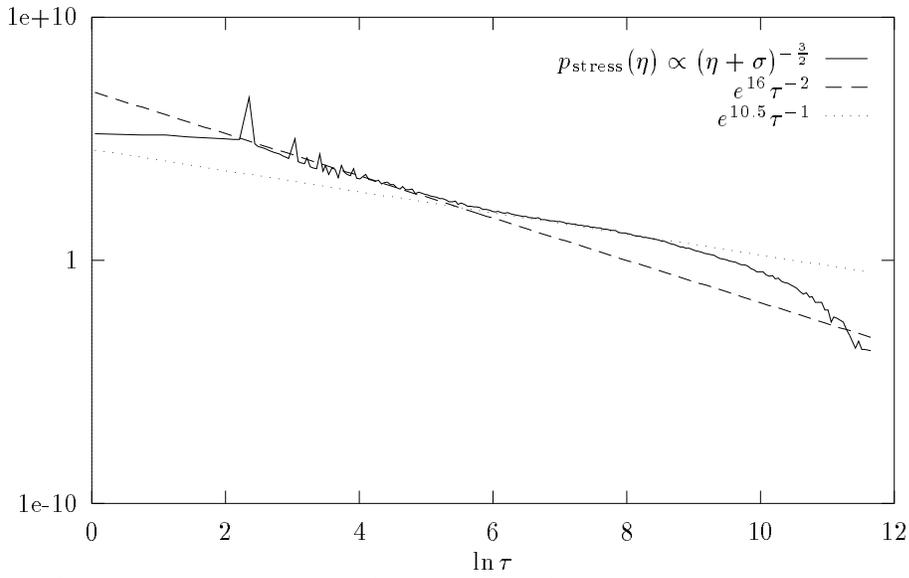}
\caption{Distribution of species lifetimes in generalised Newman
model, with $p_{\rm stress}(\eta)\propto(\eta+\sigma)^{-\frac32}$.}
\label{powdist}
\end{figure}

This result indicates that there are two domains, the first being when
$\tau<\frac{K_0}{K_1-K_0}$, where the lifetimes distribution is a
power law with exponent $-2$. This domain is more pronounced the
closer $K_1$ is to $K_2$, ie the closer $p_{\rm thresh}(x)$ is to
$p_{\rm stress}$. The other domain occurs when
$\tau>\frac{K_1}{K_1-K_0}$, where any power law will have an exponent
less than $-1$. In between, there will be a transition between the two
domains.  This result is not terribly strong, as the inequality can
also be satisfied by any distribution falling off faster than a power
law. However, it does contradict the results of the {\em Time Average
Approximation} theory of Sneppen and Newman\cite{Sneppen-Newman97} in
the case of the Lorentzian distribution, where a power law with
exponent 0 (i.e. a flat distribution) is predicted. Whilst a flat
distribution is manifestly rediculous, others are not. The TAA
predicts a power law of 1/3 for a power law stress distribution with
exponent -3/2. Figure \ref{powdist} shows the observed lifetime
distribution in this case, and the distribution never flattens out
more than $\tau^{-1}$.

Now lets us turn our attention to the Ecolab\verb|--| model to see if
similar relationship can be derived. In what follows, the species
index $i$ is dropped. Integrating equation
(\ref{ecolab--}) gives us:
\begin{displaymath}
n(t) = n_0e^{\int_0^t r - \eta(s)ds},
\end{displaymath}
and taking logarithms gives:
\begin{displaymath}
\ln n(t) = \int_0^t r - \eta(s)ds,
\end{displaymath}
since $n_0=1$ for all new species.

For the species to become extinct after time $t=\tau$, we 
require:
\begin{equation}
\int_0^t r - \eta(s)ds > \ln\sigma,\,\,\, \forall t<\tau
\end{equation}
Since time is discrete in this model, $\eta(s)$ is a piecewise
constant function, therefore the integral can be replaced by a sum so that
\begin{equation}\label{cond-eta}
\sum_{i=0}^{t-1}\eta_i < rt - \ln\sigma, \,\,\, \forall t<\tau
\end{equation}

Now inequality (\ref{cond-eta}) defines a set ${\cal
  M}\subset\reals^\tau$, and the probability of a species having
lifetime greater than $\tau$ if its reproduction rate is $r$ is given
by:
\begin{equation}
p(>\tau|r) = \int_{\cal M}\prod_{i=0}^{\tau-1} p_{\rm stress}(\eta_i)
d\eta_0 d\eta_1 \cdots d\eta_{\tau-1}
\end{equation}

Lets us first deal with sufficient conditions for inequality
(\ref{cond-eta}) to be satisfied, which are:
\begin{eqnarray}
\eta_i &<& r-\ln\sigma/\tau, \, \forall i\leq\tau\\
       &<& r, \, {\rm as }\, \sigma<1
\end{eqnarray}
Therefore a lower bound for $p(>\tau|r)$ is
\begin{equation}
p(>\tau|r) \geq
\left[\int_{-\infty}^rp_{\rm stress}(\eta)d\eta\right]^\tau.
\end{equation}

Now consider the following relation:
\begin{displaymath}
n(t+1)=(1+r-\eta_t)n(t)
\end{displaymath}
For the species not to go extinct before $t=\tau$, we require
$\eta_t<1+r, \, \forall t\leq\tau$. Therefore,
\begin{equation}
p(>\tau|r)\leq
\left[\int_{-\infty}^{(r+1)}p_{\rm stress}(\eta)d\eta\right]^\tau.
\end{equation}

Now find constants $K_0$ and $K_1$ so that
\begin{eqnarray}
K_0 p_{\rm stress}(r) \leq p_r(r) && \forall r < r_c, \exists r_c \nonumber\\
 &\leq& K_1 p_{\rm stress}(r+1) 
\end{eqnarray}
where $p_r(r)$ is the probability distribution of reproduction rates.
Since $p(>\tau)=\int p_r(r)p(>\tau|r)dr$, we find:
\begin{eqnarray}
K_0\int^{r_c} p_{\rm stress}(r)
\left[\int_{-\infty}^rp_{\rm stress}(\eta)d\eta\right]^\tau dr 
&\leq p(>\tau) \leq&
K_1\int p_{\rm stress}(r+1)
\left[\int_{-\infty}^{(r+1)}p_{\rm stress}(\eta)d\eta\right]^\tau dr\nonumber\\
\frac{K_0\rho_c^\tau}{\tau}&\leq p(>\tau)\leq&\frac{K_1}{\tau}
\end{eqnarray}

Now since $p(\tau) = p(>\tau)-p(>\tau+1)$,
\begin{eqnarray}
\frac{K_0\rho_c^\tau}{\tau}-\frac{K_1}{\tau+1}
&\leq p(\tau)\leq&
\frac{K_1}{\tau}-\frac{K_0\rho_c^\tau}{\tau+1}\\
\frac{(K_0\rho_c^\tau-K_1)\tau+K_0\rho_c^\tau}{\tau(\tau+1)}
&\leq p(\tau)\leq&
\frac{(K_1-K_0\rho_c^\tau)\tau+K_1}{\tau(\tau+1)}
\end{eqnarray}

Again, like the Newman model, we have two domains of power law
possible, an inner domain where the power law is -2, and an outer
domain where any power law is capped by -1. 
This is what is seen in Figure \ref{log-log}.

\section{Conclusion}

The Newman model owes its power law behaviour to much the same mechanism as
does Ecolab, although the assumption of constant species number hides
essential interspecies connections. Both models demonstrate a power
law exponent near $-2$ at small time scales, agreeing with the fossil
record (after Sneppen et. al \cite{Sneppen-etal95}), turning over into
a gentler power law with exponent less than -1 at larger times.

\section{Acknowledgements}

The author wishes to thank the New South Wales Centre for Parallel
Computing for use of their facilities to perform the computations for
this paper. He also wishes to thank Mark Newman for comments on this
paper.


\end{document}